# Microwave Residual Surface Resistance of Superconductors*

## Mario Rabinowitz

*Stanford Linear Accelerator Center, Stanford University, Stanford, CA 94305*
E-mail:  Mario 715@earthlink.net

**Abstract**

Two distinct models account for the microwave residual surface resistance of superconducting cavities with equally good agreement with the measured temperature and frequency dependence.  In presenting his phonon-generation model, Passow claimed that Rabinowitz' fluxoid power-loss model of residual resistance does not fit the experimental data, whereas his does.  In fact, the two models have essentially the same temperature and frequency dependence.  Furthermore, Passow's phonon-generation model cannot explain the observed sensitivity to details of sample preparation and history, while the fluxoid model can.

An analysis presented by Passow [1] showed that phonons generated in a superconductor by incident electromagnetic radiation result in a residual power loss with an equivalent surface resistance.  His expression for surface resistance is

$$R = \frac{4\pi}{c^2} \omega^2 \Lambda \left( \rho + \frac{\beta v_s \Lambda^3}{\left(1 + \omega^2 \Lambda^2 / v_s^2\right)^2} \right). \tag{1}$$

The first term represents the superconducting surface resistance derived from the BCS theory. [2, 3]  The second term is related to the power loss as electro-magnetic energy is transformed into acoustical energy.  Passow claims that this latter term becomes dominant in superconductors at low temperatures, and that it "can account for the whole of the low-temperature surface resistance measured in the purest currently available materials."  He goes on further to say, "Rabinowitz has tried to explain the residual surface rf resistance in terms of frozen-in magnetic flux. [4].  However, experiments with cavities in high magnetic fields are reported to show a different frequency dependence from that predicted by his treatment." [5, 6]

The oscillating-fluxoid power loss occurs in addition to the well-known BCS superconducting loss [2, 3], and dominates over it at low temperature. The superconducting loss decreases rapidly with decreasing temperature at low temperature, whereas the fluxoid loss has a negligible temperature dependence at low temperature in agreement with experimental observations. The effective resistivity of an oscillating fluxoid is [4, 7]

$$\rho = \left\{ \omega^2 \phi^2 H H_o \mu^2 \Big/ \left[ \rho_n^2 \left( \omega^2 M - p \right)^2 + \omega^2 \phi^2 H_o^2 \mu^2 \right] \right\} \rho_n , \qquad (2)$$

and the equivalent surface resistance is $R_f = \rho / 2\lambda$.

$\rho$ has a different meaning here than in Eq. (1), but since we are only interested in comparing the frequency dependence of Eq. (2) with that of the second term in Eq. (1), it is sufficient to retain only the common symbol $\omega$ for the angular frequency. Hence we may write the second term of Eq. (1) as

$$R_a = \frac{a\omega^2}{\left(1 + b\omega^2\right)^2} = \frac{a\omega^2}{b^2\omega^4 + 2b\omega^2 + 1} \qquad (3)$$

for the frequency dependence of the phonon contribution, where
$a = 4\pi \Lambda^4 v_s \beta / c^2$
and $b = \left(\Lambda / v_s\right)^2$. The frequency dependence of the fluxoid contribution to the surface resistance may similarly be written from Eq. (2) as

$$R_f = \frac{a_2 \omega^2}{b_2^2 \omega^4 + d\omega^2 + 1} = \frac{a_2 \omega^2}{\left(1 + b_2 \omega^2\right)^2} \text{ for } d = 2b_2 , \qquad (4)$$

where

$$a_2 = \phi^2 H H_o \mu^2 / 2\lambda p^2 \rho_n , \quad b_2 = M / p ,$$

and

$$d = \left( \phi^2 H_o^2 \mu^2 - 2Mp\rho_n^2 \right) / \left( p\rho_n \right)^2 .$$

Thus we see that the two surface resistances $R_a$ and $R_f$ have virtually the same form of frequency dependence, which become identically the same form when $\left(\phi^2 H_o^2 \mu^2 - 2Mp\rho_n^2\right)/p\rho_n^2 = 2M$. The frequency dependence of the oscillating fluxoid, $R_f$, is a little more general and includes that of $R_a$ as a special case. Hence, if $R_a$ fits the experimental data extremely well [1], the same thing may be said for $R_f$.

Although the residual loss due to acoustic coupling certainly represents a fundamental limitation and fits the temperature and frequency dependence of the experimental data quite well, one may question the magnitude of this effect. Since the fluxoid loss (and potentially other losses) fit the data equally well, the good fit itself does not point to the correctness of the one model over the other in explaining the present limits on residual resistance $R_r$. There are, however, some experimental observations which raise some doubt as to whether the acoustic loss presently represents the dominant contribution to $R_r$.

Two cavities made of the same high-purity material, having the same processing history, and as far as could be ascertained, the same bulk and surface properties, can differ by over an order of magnitude in their residual resistance. It is hard to believe that their acoustic properties differ by this much. Exposure of a high-Q cavity to CO and/or $CO_2$ (as was first done at the Stanford Linear Accelerator Center) can increase $R_r$ by more than two orders of magnitude. Again, it is unlikely that this would change the acoustic coupling to the electromagnetic radiation by this amount. In many cavities, $R_r$ first decreases with increasing field level, before it starts to increase. This seems counter to the expected increase in acoustic loss. The surface resistance can differ substantially, depending on the method of cavity cool down and on the ambient magnetic field. Again, this would not be expected from the acoustic-loss theory.

On the other hand, many, if not all, of these observations are consistent with the fluxoid power-loss model. Perhaps equally important is the fact that the fluxoid model also predicts the frequency dependence of the magnetic breakdown field $H_p'$, in agreement with experiment [8], whereas the phonon model makes no prediction. Though the observed $R_r$ and $H_p'$ are amenable to explanation by the fluxoid model, it is important to bear in mind the complexity of the phenomena and other alternative explanations may also fit the data.

## References

*Work supported by the U.S. Atomic Energy Commission.

1. C. Passow, Phys. Rev. Lett. **28**, 427 (1972).
2. D. C. Mattis and J. Bardeen, Phys. Rev. **111**, 412 (1958).
3. A.A. Abriksov, L.P. Gorkov, and I. M. Khalatnikov, Zh. Eksp. Teor. Fiz. **35**, 265 (1958) [Sov. Phys. -JETP **8**, 182 (1959).
4. M. Rabinowitz, J. Appl. Phys. **42**, 88 (1971).
5. C. Passow (private communication). Passow acted upon a remark by Kneisel et al., (Ref. 6) who conducted their experiments to investigate the influence of frozen flux on the surface resistance.
6. P. Kneisel, O. Stotltz, and J. Halbritter, IEEE Trans. Nucl. Sci. **18**, 158 (1971).
7. M. Rabinowitz, Nuovo Cimento Lett. **4**, 549 (1970).
8. M. Rabinowitz, Appl. Phys.Lett. **19**, 73 (1971).